\documentclass[twocolumn,PRA,preprintnumbers,superscriptaddress,amsmath]{revtex4}
\usepackage[latin9]{inputenc}
\usepackage{amsmath}
\usepackage{amssymb}
\usepackage{graphicx}
\usepackage[unicode=true,pdfusetitle,
 bookmarks=true,bookmarksnumbered=false,bookmarksopen=false,
 breaklinks=false,pdfborder={0 0 1},backref=false,colorlinks=false]
 {hyperref}
\hypersetup{
 colorlinks,linkcolor=red,citecolor=blue}

\makeatletter
\@ifundefined{textcolor}{}
{%
 \definecolor{BLACK}{gray}{0}
 \definecolor{WHITE}{gray}{1}
 \definecolor{RED}{rgb}{1,0,0}
 \definecolor{GREEN}{rgb}{0,1,0}
 \definecolor{BLUE}{rgb}{0,0,1}
 \definecolor{CYAN}{cmyk}{1,0,0,0}
 \definecolor{MAGENTA}{cmyk}{0,1,0,0}
 \definecolor{YELLOW}{cmyk}{0,0,1,0}
}

\makeatother

\begin{document}

\title{All-Optical control of linear and nonlinear energy transfer via Zeno
effect}

\author{Xiang Guo}

\address{Department of Electrical Engineering, Yale University, New Haven,
Connecticut 06511, USA}

\author{Chang-Ling Zou}

\address{Department of Electrical Engineering, Yale University, New Haven,
Connecticut 06511, USA}

\address{Department of Applied Physics, Yale University, New Haven, Connecticut
06511, USA}

\address{Key Laboratory of Quantum Information, University of Science and
Technology of China, Hefei, Anhui 230026, China}

\author{Liang Jiang}

\address{Department of Applied Physics, Yale University, New Haven, Connecticut
06511, USA}

\author{Hong X. Tang}

\address{Department of Electrical Engineering, Yale University, New Haven,
Connecticut 06511, USA}
\begin{abstract}
Microresonator-based nonlinear processes are fundamental to applications
including microcomb generation, parametric frequency conversion, and
harmonics generation. While nonlinear processes involving either second-
($\chi^{(2)}$) or third- ($\chi^{(3)}$) order nonlinearity have
been extensively studied, the interaction between these two basic
nonlinear processes has seldom been reported. In this paper we demonstrate
a coherent interplay between second- and third- order nonlinear processes.
The parametric ($\chi^{(2)})$ coupling to a lossy ancillary mode
shortens the lifetime of the target photonic mode and suppresses its
density of states, preventing the photon emissions into the target
photonic mode via Zeno effect. Such effect is then used to control
the stimulated four-wave mixing process and realize a suppression
ratio of $34.5$.
\end{abstract}
\maketitle
\emph{Introduction.-} The ancient Zeno's arrow paradox describes that
a flying arrow seems to be not moving if it is instantly observed.
Such concept is generalized to quantum Zeno effect (QZE) by Sudarshan
et al. \cite{Misra1977}, which states that the evolution of a quantum
system can be freezed by the frequent measurement. In theory, the
measurement of a system will have effect on the dynamics of that system,
shifting the effective energy level of a quantum system or changing
its decay rate. Zeno effects inherently result from such measurement
back-action. Apart from the frequent measurement, Zeno effect can
be equivalently realized by a continuous strong coupling to an ancillary
system, which has been proved theoretically and experimentally \cite{Kofman2001,Facchi2004,Schafer2014}.
Such effects have been widely applied in quantum system control and
engineering, including quantum state preparation \cite{Milburn1988,Itano1990,Kofman1996,Raimond2010,Signoles2014,Barontini2015,Bretheau2015},
entanglement generation \cite{Barontini2015}, autonomous quantum
error correction \cite{Erez2004,Paz-Silva2012}, and even counter-factual
quantum communication \cite{Cao2017}.

In recent years, cavity nonlinear optical processes have been widely
explored due to their important applications on Kerr comb generation
\cite{DelHaye2007,Herr2013,Xue2015,Yang2016a,Yu2016a,Guo2017}, frequency
conversion \cite{Huang1992,DeGreve2012,Guo2016a,Li2016}, new wavelength
generation \cite{Spillane2002,Carmon2007,Furst2010,Guo2016}, and
correlated photon pair generation \cite{Clemmen2009,Silverstone2014,Guo2016b}.
As an analogue of a quantum system, a photonic mode in the micro-cavity
can be treated as an energy level that allows multiple excitation.
Therefore, we could expect the similar engineering and control of
photonic mode by the Zeno effect \cite{Zou2013,Ma2014}. It is proposed
and demonstrated that the optical switching based on Zeno effect can
be realized by cavity nonlinear optical effects \cite{Wen2012,McCusker2013,Hendrickson2013,Sun2013,Chen2017}.
However, only one optical nonlinear process is involved in previous
experiments and only the linear energy transfer in the system is controlled
by the Zeno effect. 

In this Letter, we demonstrate a coherent interaction between second-
($\chi^{(2)}$) and third- ($\chi^{(3)}$) order nonlinear process
in an aluminum nitride (AlN) microring resonator. We show that not
only the linear coupling, but also the nonlinear frequency conversion
process, can be greatly suppressed by Zeno effect. Specifically, we
utilize the strong second-order optical nonlinearity in AlN microring
resonator to coherently couple the target high-$Q$ photonic mode
to a low-$Q$ ancillary mode. The coherent coupling to a lossy ancillary
mode shortens the lifetime of the target photonic mode and suppresses
its density of states, preventing the photon emissions into the target
photonic mode via Zeno effect. We first verify the Zeno effect by
probing the suppressed linear energy transfer between the external
waveguide and the micro-cavity. Such Zeno effect is then used to control
the nonlinear frequency conversion process inside the cavity, where
the photon generated in the target mode through the stimulated four-wave
mixing (FWM) process is suppressed by a factor of 34.5.

\begin{figure}[tp]
\includegraphics[width=88mm]{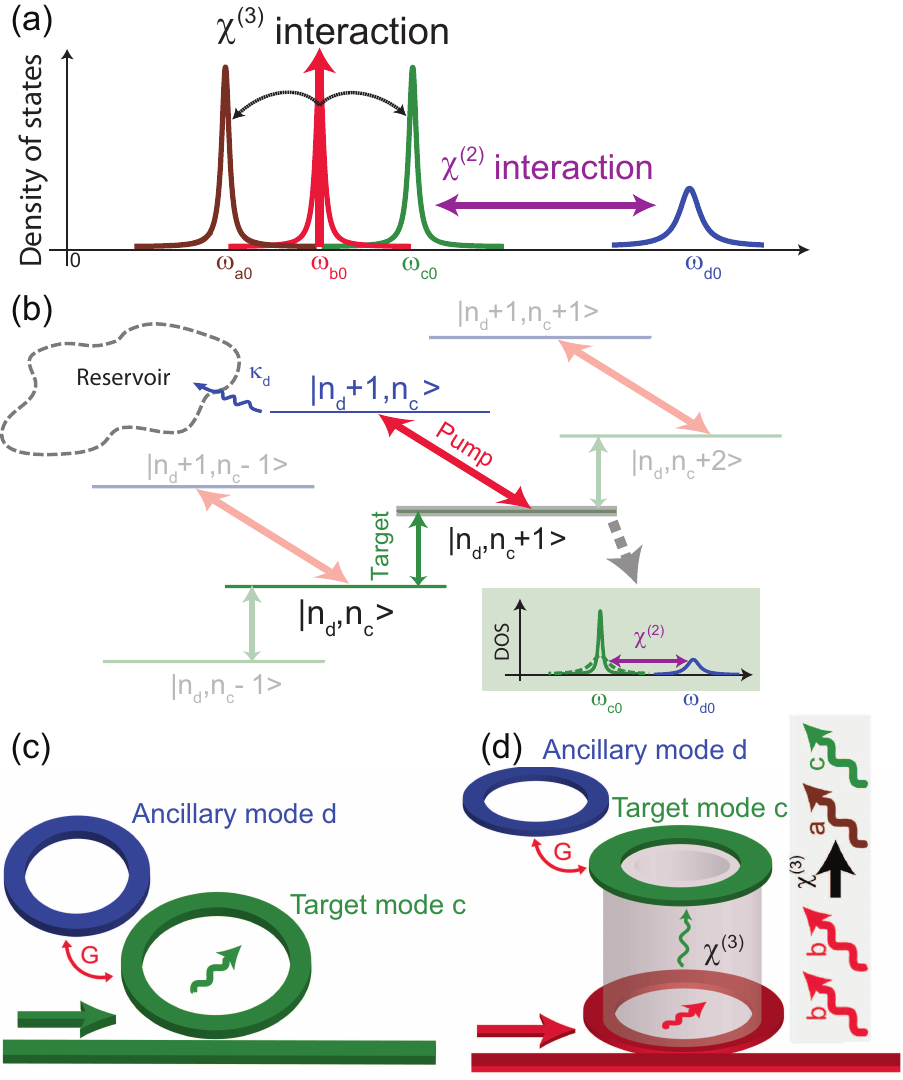}

\caption{Schematic and the energy diagram of optical modes coupling through
$\chi^{(2)}$ and $\chi^{(3)}$ nonlinearity. (a) The linear density
of states of the system. Optical modes $a$, $b$, and $c$ are coupled
through $\chi^{(3)}$ process while $b$, $c$, and $d$ are coupled
through $\chi^{(2)}$ process. (b) The energy diagram describing the
nonlinear coupling between the target photonic mode $c$ and the ancillary
photonic mode $d$ via three-wave mixing process. $n_{c}$($n_{d}$)
stands for the cavity photon number for mode $c$ ($d$). Inset: the
dramatically suppressed density of states for the target mode when
coherently coupled to the ancillary mode. (c) Probing the Zeno effect
with linear waveguide-to-microring coupling scheme. (d) Controlling
nonlinear wavelength conversion process the via Zeno effect. The photon
emissions into the targeted optical mode via four-wave mixing process
are suppressed when the target optical mode is coupled to the dissipative
ancillary mode. }
\label{Fig1}
\end{figure}

\emph{System and model.- }As a material possessing both $\chi^{(2)}$
and $\chi^{(3)}$ nonlinearity \cite{Jung2014}, AlN has been used
to realize high-efficiency second-harmonic generation \cite{Guo2016}
(using $\chi^{(2)}$) and third-harmonic generation \cite{Surya2018}
(using $\chi^{(3)}$). It is worth noting that simultaneous generation
of both second- and third- harmonics has also been observed in the
case of extreme nonlinear optics \cite{Tritschler2003}, where both
nonlinear optical processes originate from $\chi^{(3)}$ nonlinearity.\emph{
}Our experimental system contains a waveguide coupled AlN microring
resonator \cite{Guo2016,Guo2016a}, which supports a variety of optical
modes of different wavelengths. In the following study, we focus on
three optical modes located in the telecom band ($a$, $b$, $c$)
and one optical mode in the visible band ($d$). When the microring
is cold, i.e. not pumped by any laser, the density of states of all
these optical modes are shown in Fig.$\,$\ref{Fig1}(a). The telecom
optical modes ($a$, $b$, $c$) have a higher and narrower density
of states because of their longer lifetime ($\thicksim210\,\mathrm{ps}$)
compared to the visible mode $d$ ($\thicksim45\,\mathrm{ps}$). The
third-order nonlinear  interaction leads to FWM process where two
photons in mode $b$ are converted to a pair of photons in mode $a$
and $c$ ($b+b\rightarrow a+c$), as illustrated in Fig.$\,$\ref{Fig1}(a).
Due to the second-order nonlinear  effect, modes $b$, $c$ are also
participating in the three-wave mixing (TWM) process ($b+c\rightarrow d$)
which involves the high-loss visible mode $d$. Under a pump at mode
$b$, the system can be described by the Hamiltonian \cite{SupplyM}
\begin{eqnarray}
\frac{H}{\hbar} & = & \sum_{o\in\{a,b,c,d\}}\omega_{o,0}\hat{o}^{\dagger}\hat{o}+g_{2}\left(\hat{b}^{\dagger}\hat{c}^{\dagger}\hat{d}+c.c.\right)\nonumber \\
 &  & +g_{3}\left[\left(\hat{b}^{\dagger}\right)^{2}\hat{a}\hat{c}+c.c.\right]+\epsilon_{b}\left(i\hat{b}^{\dagger}e^{-i\omega_{b}t}+c.c.\right).
\end{eqnarray}
Here, $\hat{o}$ is the Bosonic operator for mode $o$ with $o\in\{a,b,c,d\}$,
and $\omega_{o,0}$ is the angular frequency of mode $o$. The intrinsic,
external and total loss rate of mode $o$ are denoted as $\kappa_{o,0}$,
$\kappa_{o,1}$ and $\kappa_{o}=\kappa_{o,0}+\kappa_{o,1}$, respectively.
The input pump field $\epsilon_{b}=\sqrt{2\kappa_{b,1}P_{b}/\hbar\omega_{b}}$,
where $P_{b}$ and $\omega_{b}$ are the power and frequency of the
pump laser, respectively. $g_{2}$ and $g_{3}$ are the single-photon
nonlinear coupling strength for the TWM and FWM processes, respectively.
The numerical values for all the relevant parameters are summarized
in the supplementary materials.

Under a strong coherent pump field in mode $b$, the photons in mode
$c$ can be coherently converted to the high-loss mode $d$ and get
dissipated to the environment (Fig.$\,$\ref{Fig1}(b)). For the sake
of convenience, in the following context we name the long-lifetime
mode $c$ as the ``target'' photonic mode and the short-lifetime
mode $d$ as the ``ancillary'' mode. This coherent coupling between
the target mode $c$ and the ancillary mode $d$ can be regarded as
a continuous measurement of mode $c$. And the backaction of such
measurement leads to a counter-factual result that the ability for
the photons to couple into the target mode $c$ is suppressed. This
is an analogue to the Zeno effect where frequent measurement inhibits
the occupation of an energy level. Alternatively, we can understand
the Zeno effect as the reduction of the density of states, as shown
in the inset of Fig.$\,$\ref{Fig1}(b), due to coupling with the
dissipative ancillary mode.

\begin{figure}[tp]
\includegraphics[width=8.8cm]{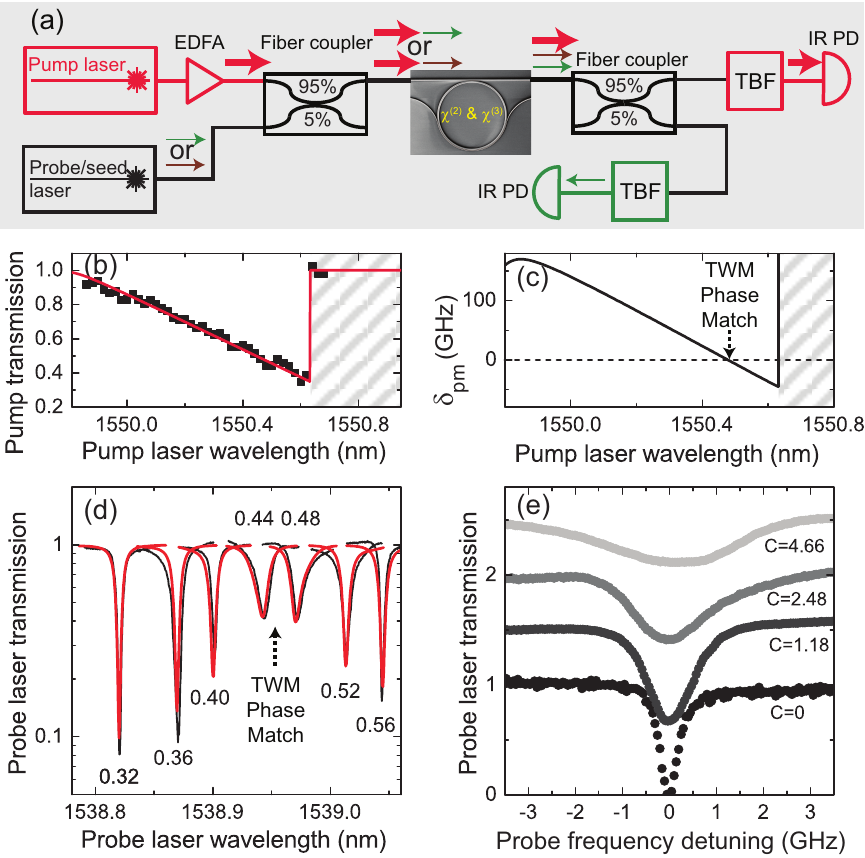}

\caption{Probing the Zeno effect with linear coupling scheme. (a) Experimental
setup. A strong laser is used to coherently drive the optical mode
$b$. A probe laser can either directly probe the Zeno effect using
linear coupling scheme by scanning across the resonance of target
photonic mode $c$ or using nonlinear scheme by scanning across the
resonance of mode $a$. TBF: tunable bandpass filter. (b) The transmitted
power of the strong pump laser when scanning across the resonance
of mode $b$. A triangle-like shape is observed due to the thermal
effect. (c) The tuning of phase mismatch $\delta_{pm}$ during the
pump laser wavelength scan. At a certain pump wavelength the phase-match
condition is fulfilled. (d) The measured transmission spectra of the
target photonic mode $c$ under different pump laser wavelength. The
pump laser wavelength ($\lambda_{b}-1550\,\mathrm{nm}$) is labeled
near each curve. When the phase mismatch gradually approaches 0, a
much lower-extinction resonance dip is observed. The black curves
are experimental data, which agree with the red theoretical curves.
(e) The measured transmission of the target photonic mode $c$ with
phase-matching condition $\delta_{pm}=0$ satisfied. With larger pump
laser photon number (hence larger cooperativity $C$), the mode $c$'s
extinction is reduced and the Zeno effect is more prominent. }
\label{Fig2}
\end{figure}

\emph{Linear probe of the Zeno effect.- }To experimentally verify
the Zeno effect induced by the TWM, we directly probe the linear energy
transfer from the external waveguide to the target mode $c$, which
has been employed in previous related works \cite{Wen2012,McCusker2013,Hendrickson2013,Sun2013,Chen2017}.
This linear probe scheme is shown in Fig.$\,$\ref{Fig1}(c), where
a bus waveguide is used to send photons into the target photonic mode
through the evanescent coupling between waveguide and the microring. 

When the dissipative ancillary mode is parametrically coupled with
the target mode via the TWM process, the photons remain in the bus
waveguide with suppressed coupling into the microring. Quantitatively,
when scanning the frequency of the probe laser $\left(\omega_{c}\right)$
across the resonance of the target mode $c$, the transmission spectrum
can be calculated as \cite{SupplyM}
\begin{eqnarray}
t_{c} & = & \left|1+\frac{2\kappa_{c,1}}{-i\delta_{c}-\kappa_{c}+\frac{g_{2}^{2}\left|\beta\right|^{2}}{(-i\delta_{d}-\kappa_{d})}}\right|^{2}.\label{eq:ZenoLinear}
\end{eqnarray}
Here, $\delta_{c}=\omega_{c0}-\omega_{c}$ and $\delta_{d}=\omega_{d0}-\omega_{b}-\omega_{c}$
are the frequency detunings for mode $c$ and $d$, respectively,
which are expressed in terms of the pump frequency $\omega_{b}$ and
probe frequency $\omega_{c}$. $\beta=\epsilon_{b}/\left[-i\left(\omega_{b,0}-\omega_{b}\right)-\kappa_{b}\right]$
represents the pump mode field under the non-depletion approximation.
Compared with the transmission spectrum of $c$ without the pump ($\beta=0$)
that $t_{c}=\left|1+2\kappa_{c,1}/(-i\delta_{c}-\kappa_{c})\right|^{2}$,
a maximum modification of the target mode's transmission will happen
when the phase-matching condition for the TWM process is fulfilled,
i.e $\delta_{pm}\triangleq\omega_{d0}-\left(\omega_{b}+\omega_{c0}\right)=0$,
which physically mean that mode $c$ (with a resonant frequency $\omega_{c0}$)
and $d$ (with a resonant frequency $\omega_{d0}$) are resonant coupled
to each other in the rotating frame of the pump laser frequency $\omega_{b}$.
When phase matched, the coherent coupling to the ancillary mode $d$
introduces an extra loss term $\gamma=g_{2}^{2}\left|\beta\right|^{2}/\kappa_{d}$
to $\kappa_{c}$. When the coupling to the ancillary mode is very
strong, i.e. $\gamma\gg\kappa_{c}$, the coherent coupling will quickly
dissipate any photons that couple into the target photonic mode $c$.
This leads to a counter-factual result that the ability for the photons
to couple into the target mode $c$ is suppressed and results in a
much larger transmission $t_{c}$, as we will show next. 

The experimental setup is illustrated in Fig.$\,$\ref{Fig2}(a).
One pump laser is used to coherently drive the optical mode $b$,
acting as the pump for the TWM process. The on-chip power of the pump
laser varies from $0$ to around $180\,\mathrm{mW}$. In experiment,
we scan the frequency of pump laser $\omega_{b}$ across the resonance
of the pump mode $b$. Since the value of $\delta_{pm}$ is very important
for observing the Zeno effect, we first explain the tuning of $\delta_{pm}$
during the pump laser scan.\textbf{ }Due to the thermal effects, a
frequency red shift will happen when scanning the pump laser, which
results in an extended transmission spectrum with a triangle shape
\cite{Carmon2004}. Figure \ref{Fig2}(b) shows a typical pump laser
transmission spectrum. The thermal effects induced by the pump would
also affect the other resonances in the same microring cavity, albeit
at different rate. As a result, the value of $\delta_{pm}$ will also
change during the pump laser scanning process (Fig. \ref{Fig2}(c)).
It is indicated that at a certain pump wavelength, the phase-matching
condition $\delta_{pm}=0$ is satisfied. In this condition, the target
mode $c$ will be \emph{resonantly} coupled to the ancillary mode
$d$ and the effect of TWM will be prominent. 

In the experiment, we measure the transmission spectrum of the target
mode $c$ during each scanning step of pump laser $\omega_{b}$. Each
step corresponds to a different phase mismatch $\delta_{pm}$ and
pump laser field $\beta$. The on-chip probe laser power was set to
be $0.1\,\mathrm{mW}$, much weaker than the pump laser power. In
the measured transmission spectra of the target mode $c$ (Fig. \ref{Fig2}(d)),
we can see that originally the transmission is of normal Lorentzian
shape, which corresponds to the situation where the TWM is not prominent
due to an unsatisfied phase-matching condition ($\delta_{pm}\gg\kappa_{c}$).
The photons from bus waveguide can be efficiently coupled to the target
photonic mode, leading to a high-extinction dip in the transmission
spectrum. At a certain point, when the phase mismatch $\delta_{pm}=0$
and $\beta$ is considerably large, a broadened transmission spectrum
of the target mode $c$ is observed. Noticeably, the much shallower
extinction dip manifests itself as an indication of Zeno effect, with
suppressed coupling between the bus waveguide and the microring. As
the the pump laser wavelength changes further, the coupling to the
ancillary mode gets lost due to unsatisfied phase-matching condition
($\delta_{pm}\neq0$) and the spectrum of target mode $c$ returns
to the high-extinction Lorentzian shape. We observe in Fig. \ref{Fig2}(d)
that the simulated transmission spectra (red) of mode $c$ matches
very well with the measured curve (black), indicating a good theoretical
understanding of this process. By increasing the power of the pump
laser, the cooperativity \cite{Guo2016a} $C=\frac{g_{2}^{2}\left|\beta\right|^{2}}{\kappa_{c}\kappa_{d}}$
of the TWM process will increase proportionally. We can therefore
observe the suppression of the linear coupling with increasing fidelity,
as shown in Fig.$\,$\ref{Fig2}(e).

\begin{figure}[!t]
\includegraphics[width=8.8cm]{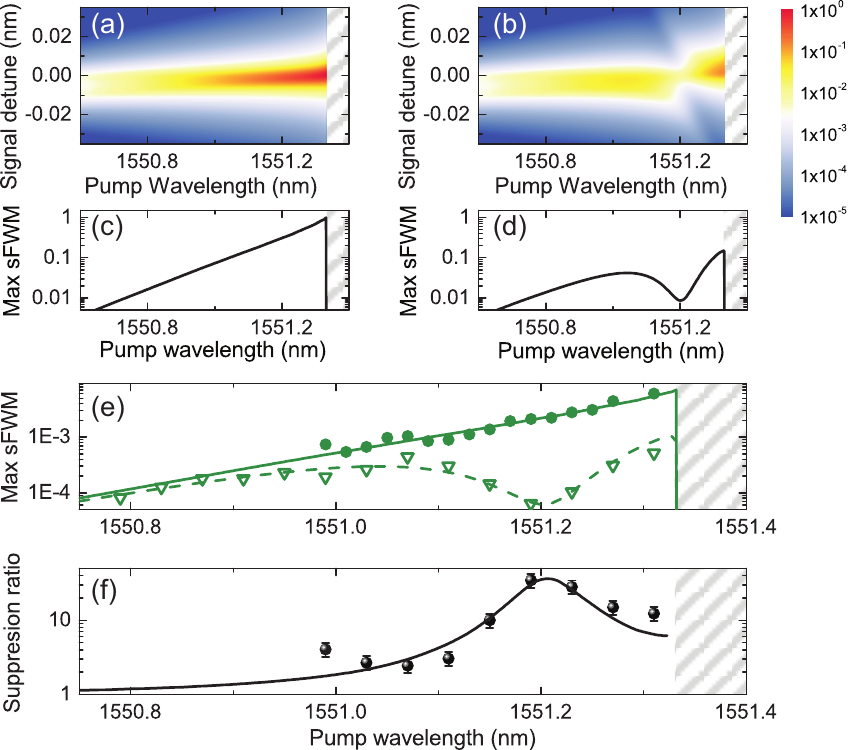}

\caption{Numerical simulation and experimental measurement of the suppressed
four-wave mixing efficiency due to Zeno effect. (a-b) The intensity
of stimulated four-wave mixing in target mode $c$ (a) with or (b)
without $\chi^{(2)}$ process. The signal laser is scanning across
mode $a$ while the pump laser is scanning across mode $b$. (c-d)
The maximum stimulated four-wave mixing intensity as the pump laser
scanning across mode $b$ when $\chi^{(2)}$ interaction (c) does
not show up (d) shows up. (e) Experimental result of the stimulated
four-wave mixing efficiency with and without $\chi^{(2)}$ interaction.
The solid circles (without $\chi^{(2)}$) and open triangles (with
$\chi^{(2)}$) represent the experimental measurements while the solid
(without $\chi^{(2)}$) and dashed (with $\chi^{(2)}$) lines corresponds
to theoretical fitting. (f) The suppression ratio of four-wave mixing
efficiency as a function of pump wavelength.}

\label{Fig3}
\end{figure}

\emph{Controlling FWM via Zeno effect.- }As the Zeno effect by TWM
is confirmed by measuring the linear energy transfer in the system,
we then investigate how the Zeno effect can be used to control the
nonlinear processes in the cavity. Specifically, we study the suppression
of photon generation in the target mode through the stimulated FWM,
which is a nonlinear energy transfer process, as depicted in Fig.$\,$\ref{Fig1}(d).
By controlling the temperature of the photonic chip, we can determine
whether the phase-matching condition of the TWM process can be fulfilled
\cite{Guo2016}. As we show next, by switching on and off the TWM
process, we can control the FWM efficiency by more than one order
of magnitude. 

At certain temperature, the phase-matching condition is not fulfilled,
i.e. $g_{2}^{2}\left|\beta\right|^{2}/\delta_{d}\ll\kappa_{c}$, and
TWM's effect is negligible. In this case, we can study the pure FWM
process, i.e. ignoring the $\chi^{(2)}$ related term in the system's
Hamiltonian. As mode $b$ is strongly pumped by a pump laser with
frequency $\omega_{b}$, we stimulate the FWM process by sending a
seed laser around mode $a$ and record the generated light intensity
in the target mode $c$. The nonlinear energy transfer to the target
mode $c$ through the FWM process can be calculated by \cite{SupplyM}
\begin{eqnarray}
P_{\mathrm{c,out}} & = & \frac{2\kappa_{a,1}}{\delta_{a}^{2}+\kappa_{a}^{2}}\frac{2\kappa_{c,1}}{\delta_{c}^{2}+\kappa_{c}^{2}}\frac{\omega_{c}}{\omega_{a}}\left|\beta^{4}\right|g_{3}^{2}P_{\mathrm{a,in}},\label{eq:FWM}
\end{eqnarray}
 where $P_{\mathrm{a,in}}$ is the input power of the seed laser around
mode $a$. During the scanning of pump laser $\omega_{b}$, the pump
laser frequency gradually approach the resonance of mode $b$ and
suddenly jump out of resonance due to thermal bistability, which is
already explained and shown in Fig.$\,$\ref{Fig2}(b). By considering
the thermal induced frequency shift into Eq.$\,$(\ref{eq:FWM}),
the normalized efficiency of the stimulated FWM as a function of both
pump laser wavelength and seed laser wavelength can be estimated,
as shown in Fig.$\,$\ref{Fig3}(a). By extracting the maximum stimulated
FWM efficiency at each pump wavelength {[}Fig.$\,$\ref{Fig3}(c){]},
we observe that the photon emissions in the target mode $c$ through
stimulated FWM increases monotonically with the pump wavelength approaching
the resonance of mode $b$. 

By tuning the device's temperature, we can fulfill the phase-matching
condition for TWM process. We then consider the case when the TWM
exists and couples the target mode $c$ with an ancillary mode $d$.
In this case, the nonlinear energy transfer to the target mode is
modified as
\begin{eqnarray}
P_{\mathrm{c,out}}^{\prime} & = & P_{\mathrm{c,out}}/\left|1+\frac{g_{2}^{2}\left|\beta\right|^{2}}{(-i\delta_{d}-\kappa_{d})(-i\delta_{c}-\kappa_{c})}\right|^{2}.
\end{eqnarray}
The additional term $\frac{g_{2}^{2}\left|\beta\right|^{2}}{(-i\delta_{d}-\kappa_{d})(-i\delta_{c}-\kappa_{c})}$
due to coherent coupling between mode $c$ and $d$ can lead to a
dramatic decrease of the nonlinear energy transfer efficiency. The
numerical simulation of the stimulated FWM process when the TWM is
involved is shown in Fig.$\,$\ref{Fig3}(b). Compared to Fig. \ref{Fig3}(a),
a clear suppression of the efficiency is observed when the phase-matching
condition ($\delta_{pm}=0$) of TWM process is satisfied. Again we
extract the maximum stimulated FWM efficiency at each pump wavelength,
as shown in Fig.$\,$\ref{Fig3}(d). A clear dip of FWM efficiency
appears when the phase-matching condition is satisfied. This suppressed
FWM efficiency signifies the appearance of Zeno effect in the nonlinear
energy transfer process. 

The experimental results about such suppression of stimulated FWM
are summarized in Figs.$\,$\ref{Fig3}(e) and (f). When the TWM is
negligible, the measured stimulated FWM efficiency depends on the
pump wavelength as shown by the solid circles in Fig.$\,$\ref{Fig3}(e).
In contrast, when the phase-matching condition of TWM process is fulfilled,
a clear suppression of FWM efficiency is shown by the open triangles
as depicted in Fig.$\,$\ref{Fig3}(e). The solid and dashed lines
come from the theoretical fitting and the used fitting parameters
are summarized in the supplementary materials. To quantify the suppression
of nonlinear frequency conversion process induced by the Zeno effect,
we plot the suppression ratio against the pump wavelength, as shown
in Fig.$\,$\ref{Fig3}(f). It can be observed that at a certain pump
wavelength, when the phase-matching of TWM is satisfied, the highest
suppression ratio of 34.5 is achieved. This large suppression ratio
clearly demonstrates the control of the system's nonlinear dynamics
utilizing the Zeno effect. Note that the experimental result deviates
slightly from the theoretical curve at the pump wavelength close to
1551.0 nm. This could be induced by the weak back-scattering inside
the microring, which is not considered in the theoretical model.

\emph{Conclusion.- }We have investigated the all-optical control of
linear and nonlinear energy transfer in a nonlinear microring cavity
by combining the TWM and FWM processes. We show that the efficiency
of stimulated four-wave mixing is suppressed by a factor of 34.5 through
Zeno control, which is implemented by coherently coupling the target
photonic mode to the high-loss ancillary mode. The demonstrated Zeno
control on a photonic chip can be a useful tool for tailoring the
local mode's density of state and the dispersion, suppressing or enhancing
the nonlinear optics process. The ample nonlinear effects in the micro-cavity
may also lead to potential applications in quantum nonlinear optics,
such as entangled photon pairs generation \cite{Grassani2015,Silverstone2015}
and $in-situ$ frequency conversion \cite{Guo2016a,Li2016,Vernon2016}
into other wavelength bands. 
\begin{acknowledgments}
Facilities used for device fabrication were supported by Yale SEAS
cleanroom and Yale Institute for Nanoscience and Quantum Engineering.
The authors thank Michael Power and Dr. Michael Rooks for assistance
in device fabrication. We acknowledge funding support from an AFOSR
MURI grants (FA9550-15-1-0029, FA9550-15-1-0015), LPS/ARO grant (W911NF-14-1-0563),
NSF EFRI grant (EFMA-1640959), DARPA SCOUT program and the Packard
Foundation. L.J. acknowledges support from the Alfred P. Sloan Foundation
(BR2013-049) and the Packard Foundation (2013-39273).
\end{acknowledgments}

\bibliographystyle{apsrev4-1_v2}
\bibliography{Zeno2}

\end{document}